\documentclass[preprint,showpacs,preprintnumbers,amsmath,amssymb]{revtex4}

\usepackage{graphicx}% Include figure files
\usepackage{dcolumn}% Align table columns on decimal point
\usepackage{bm}% bold math
\begin{document}
\title{Topological entanglement entropy in bilayer quantum Hall systems}

\author{Myung-Hoon Chung}
\affiliation{College of Science and Technology, Hongik University,
Sejong 339-701, Korea\\
mhchung@hongik.ac.kr}

\date{\today}

\begin{abstract}
We calculate the topological entanglement entropy in bilayer
quantum Hall systems, dividing the set of quantum numbers into
four parts. This topological entanglement entropy allows us to
draw a phase diagram in the parameter space of layer separation
and tunneling amplitude. We perform the finite size scaling
analysis of the topological entanglement entropy in order to see
the quantum phase transition clearly.
\end{abstract}


\maketitle

\section{Introduction}

None of local order parameters in some cases can distinguish the
types of quantum order, because quantum nature itself is nonlocal.
To overcome this difficult situation, Kitaev and Preskill
\cite{Kitaev} have introduced the topological entanglement
entropy, which is obtained by the well-designed partition and the
clever linear combination of the corresponding entanglement
entropies. It is of interest to look for an explicit microscopic
model that realizes quantum order characterized by the topological
entanglement entropy. The purpose of this paper is to present the
microscopic model related to the topological entanglement entropy.

A system with a mass gap in two spatial dimensions can exhibit
topological order \cite{Wen}. A mass gap is the key ingredient for
the incompressible quantum Hall state \cite{Laughlin}. The
quasiparticle excitations in the quantum Hall system obey
fractional statistics \cite{Arovas}. Furthermore, Haldane
\cite{Haldane} showed that, in the quantum Hall system, the
ground-state degeneracy depends on whether the geometry is either
sphere or torus. These features of mass gap, fractional
statistics, and dependence of degeneracy are all topological
properties. Hence it is natural to consider the topological
entanglement entropy in the quantum Hall system.

For applications of the topological entanglement entropy in
relation with quantum phase transitions, we consider bilayer
quantum Hall systems, where it is simpler to introduce
controllable parameters into the Hamiltonian of the system. In
bilayer quantum Hall systems, two parameters to be controlled are
the layer separation $d$ and the inter-layer tunneling amplitude
$t$. The experimental strong evidence for the quantum phase
transition of bilayer systems was a strong enhancement in the
zero-bias inter-layer tunneling conductance for a small $d$ system
at total Landau level filling factor $\nu=1$ \cite{Spielman}.
Theoretically, if $d$ goes to $\infty$ for fixed $\nu = 1$, the
bilayer system becomes a set of two single layer systems for $\nu
= \frac{1}{2}$. For a large $d$ system, the ground state would be
compressible, which is not a quantum Hall state. It is clear that
phase transition takes place at the critical value $d_c$ which is
met while $d$ changes from $0$ to $\infty$. The main concern is
now to draw the phase diagram of the system in the parameter space
of $d$ and $t$.

Pseudospin notation for the layer degree of freedom is used to
find the phase diagram, by calculating the pseudospin
magnetization \cite{Schliemann}. However this pseudospin
magnetization approach is not conclusive because calculations of
varying $d$ for fixed $t$ give different results from those of
varying $t$ for fixed $d$. Since the pseudospin is a local order
parameter, the approach of the pseudospin magnetization may not
provide perfect explanation for the nature of quantum phase
transitions in bilayer quantum Hall systems.

In this paper, we focus on the topological entanglement entropy,
which is the most natural order parameter to study phase
transition in bilayer quantum Hall systems. Using exact
diagonalization, we numerically evaluate the topological
entanglement entropy in finite size systems. We analyze the
behavior of the topological entanglement entropy as we vary $d$
for fixed $t$. We also carry out the analysis of finite size
scaling. We will show that the topological entanglement entropy
provides a different phase boundary from that of the pseudospin
magnetization in bilayer quantum Hall systems. This difference is
controversial. More detailed experimental measurements are
required to resolve the issue of topological entanglement entropy
as an order parameter for bilayer quantum Hall systems.

\section{Hamiltonian}

We start with presenting the bilayer quantum Hall system in terms
of the second quantized form of the Hamiltonian in a torus
geometry within the lowest Landau level approximation. It is known
that the Landau level degeneracy $N$ is determined by the magnetic
field strength $B$ and the square torus area $L^2$ as $L^2 =2 \pi
N l_{B}^{2}$ where $l_B = \sqrt{\hbar c/eB}$ is the magnetic
length. It is convenient to measure all distances in the unit of
$l_B$, and energies in the unit of $e^2/\epsilon l_B$, where
$\epsilon$ is a dielectric constant.

The two-body interaction between electrons is described by the
periodic Coulomb interaction $U(\vec{x}_{i}-\vec{x}_{j})$, which
is written in terms of position variables $\vec{x}$ and momentum
variables $\vec{k}$ by using the Fourier transformation:
\[
U(\vec{x}_{i}-\vec{x}_{j}) = \frac{e^{2}}{\epsilon
|\vec{x}_{i}-\vec{x}_{j}|} = \lim_{\mu \rightarrow 0}
\frac{e^{2}}{\epsilon (2\pi)^{3}}\int d^{3}k
\frac{4\pi}{k^{2}+\mu^{2}} \exp[i\vec{k}\cdot
(\vec{x}_{i}-\vec{x}_{j})],
\]
where $\mu$ is introduced in order to take into account the
infrared divergence. If two electrons are on the same (different)
layer, intra (inter) layer, the third component of
$(\vec{x}_{i}-\vec{x}_{j})$ is $0$ ($d$). In the torus geometry of
the finite size $L^{2}$, the first and second components $k_1$ and
$k_2$ out of $\vec{k}$ turn to the discrete integers $n_{1}$ and $
n_{2}$, while $k_3$ is kept as continuous. Then
$U(\vec{x}_{i}-\vec{x}_{j})$ between inter-layer electrons is
written as
\begin{eqnarray*}
U(\vec{x}_{i}-\vec{x}_{j})=\frac{e^{2}}{\epsilon
2\pi^{2}}\left(\frac{2\pi}{L}\right)^{2}\{
\lim_{\mu \rightarrow 0} \int_{-\infty}^{\infty}dk_{3}\frac{ \exp(ik_3 d)}{k_{3}^{2}+\mu^{2}} \\
+\sum_{(n_{1},n_{2})\neq (0,0)} \int_{-\infty}^{\infty} dk_{3}
\frac{\exp[i\frac{2\pi}{L}\vec{n}\cdot (\vec{x}_{i}-\vec{x}_{j}) +
ik_3 d]}
{(\frac{2\pi}{L}n_{1})^{2}+(\frac{2\pi}{L}n_{2})^{2}+k^{2}_{3}}
\},
\end{eqnarray*}
where the first term is extracted as the infrared divergence part
for the case of $n_{1}=n_{2}=0$. Integrating out $k_{3}$, we
obtain
\begin{eqnarray*}
U(\vec{x}_{i}-\vec{x}_{j})=\lim_{\mu \rightarrow 0} \frac{e^{2}}{\epsilon L} 2\pi \frac{ \exp(-\mu d)}{L\mu} \\
+\frac{e^{2}}{\epsilon L}\sum
\frac{\exp[-\sqrt{n_{1}^{2}+n_{2}^{2}}\frac{2\pi}{L}d]}
{\sqrt{n_{1}^{2}+n_{2}^{2}}} \exp[i\frac{2\pi}{L}\vec{n}\cdot
(\vec{x}_{i}-\vec{x}_{j})].
\end{eqnarray*}
Expanding $\exp(-\mu d)$ into the Taylor series with respect to
$\mu$, we rewrite the first term of $U(\vec{x}_{i}-\vec{x}_{j})$
as follows:
\begin{equation*}
\frac{e^{2}}{\epsilon L} 2\pi \frac{ \exp(-\mu d)}{L\mu} =
\frac{e^{2}}{\epsilon l_{B}} \frac{ 1 }{N l_{B} \mu} -
\frac{e^{2}}{\epsilon l_{B}} \frac{ d }{N l_{B}} +
\mathcal{O}(\mu).
\end{equation*}
The infinite first term explains the infrared divergence, and it
should be canceled by uniform positive background charge
\cite{Shibata}. The finite second term which depends on $d$
contributes to a static charging-energy \cite{Nomura}.

The Fourier transformation and handling the infrared divergence
make it straightforward to derive the second quantized Hamiltonian
by using the single particle wave-function. For the lowest Landau
level in the torus geometry, the $j$-th single particle
wave-function $\psi_{j}(x,y)$ \cite{Cristofano} is given by
\begin{eqnarray*}
\psi_{j}(x,y)=(\frac{1}{L\sqrt{\pi}l_{B}})^{1/2}
\exp(-\frac{y^{2}}{2l_{B}^{2}}) \nonumber \\
\times \sum_{k=-\infty}^{\infty}\exp[-\pi N (k+\frac{j}{N})^{2} +
i 2 \pi N (k+\frac{j}{N})(\frac{x+iy}{L})].
\end{eqnarray*}

The Hamiltonian $H$ for the bilayer quantum Hall system is the sum
of the Coulomb interaction term $H_i$ and the single particle
inter-layer tunneling term $H_t$ such as
\begin{equation}
H = H_{i} + H_{t}.
\end{equation}

Based on the above wave functions $\psi_{j}(x,y)$ and the periodic
Coulomb interaction $U(\vec{x}_{i}-\vec{x}_{j})$, we obtain the
second quantized form of $H_{i}$. The Hamiltonian is expressed in
terms of creation and annihilation operators
$c_{j\sigma}^{\dagger}$ and $c_{j\sigma}$, where pseudospin
$\sigma = \uparrow$ or $\downarrow$ is used to describe different
layers. We get
\begin{equation}
H_{i}=H_{\uparrow}+H_{\downarrow}+H_{\uparrow\downarrow}-\frac{e^{2}}{\epsilon
l_{B}}\frac{d}{l_{B}}\frac{N_{\uparrow}N_{\downarrow}}{N},
\end{equation}
where $H_{\uparrow}$ ($H_{\downarrow}$) presents the interaction
between electrons in up (down) layer, $H_{\uparrow\downarrow}$ is
the inter-layer Hamiltonian, and $N_{\uparrow}$($N_{\downarrow}$)
in charging-energy term is the number of electrons in up (down)
layer. The value of product $N_{\uparrow}N_{\downarrow}$ is
maximized at $N_{\uparrow}=N_{\downarrow}=N/2$ with the constraint
of $N_{\uparrow}+N_{\downarrow}=N$. Without the last term, all
electrons stay in a single layer according to Hund's rule for a
small $d$ as shown in Table \ref{tab:tab1}.

Following the procedure given by Yoshioka-Halperin-Lee
\cite{Yoshioka}, we find
\begin{eqnarray*}
H_{\sigma}&=&\frac{1}{2}\sum_{a,b=0}^{N-1}V(a,b;0)\sum_{k=0}^{N-1}c_{k+a\sigma}^{\dagger}
c_{k+b\sigma}^{\dagger}c_{k+a+b\sigma}c_{k\sigma},\nonumber \\
H_{\uparrow\downarrow}&=&\sum_{\sigma \not= \sigma^{\prime}}
\frac{1}{2}\sum_{a,b}V(a,b;d) \sum_{k}c_{k+a\sigma}^{\dagger}
c_{k+b\sigma^{\prime}}^{\dagger}c_{k+a+b\sigma^{\prime}}c_{k\sigma}.
\end{eqnarray*}
Here the orbital index $j$ in $c_{j\sigma}^{\dagger}$ should
satisfy the periodic condition such that
$c^{\dagger}_{j+N~\sigma}=c^{\dagger}_{j~\sigma}$. The
coefficients $V$ in the Hamiltonian are given by
\begin{eqnarray*}
V(a,b; d)=\sum_{m,l\in \mbox{Z}}g(Nm+a,l)
\exp(-\frac{\pi}{N}\{l^{2}+(Nm+a)^{2}\})\cos(\frac{2\pi}{N}lb), \\
g(i,j)=\frac{e^{2}}{\epsilon l_{B}}\frac{1}{N}
\frac{\exp[-\sqrt{\frac{2\pi}{N}(i^{2}+j^{2})}\frac{d}{l_{B}}]}
{\sqrt{\frac{2\pi}{N}(i^{2}+j^{2})}},
\end{eqnarray*}
where the case of $g(0,0)$ is excluded since the infrared
divergence was already handled \cite{Chung1}.

The single particle tunneling term $H_{t}$ is written as
\begin{eqnarray}
H_{t} = - \frac{e^{2}}{\epsilon l_{B}} t\frac{1}{2}
\sum_{j}\sum_{\sigma\sigma^{\prime}} c^{\dagger}_{j\sigma}
s^{\sigma\sigma^{\prime}}_x c_{j\sigma^{\prime}},
\end{eqnarray}
where $s^{\sigma\sigma^{\prime}}_x$ is the pseudospin Pauli matrix
and $t$ measures the inter-layer tunneling amplitude.

After we set the Hamiltonian, we perform exact diagonalization to
find the ground state of a small system. Doing consistency checks,
we show explicit values of the ground state energy for the system
of $N=8$ in Table \ref{tab:tab1}.

When the layer separation vanishes ($d=0$) and no tunneling is
allowed ($t=0$), the Hamiltonian has pseudospin SU(2) rotational
symmetry. It means $[H_{i}, \vec{S}] = 0$, where $\vec{S}$ is the
total pseudospin operator \cite{Nomura}. Then the ground state
should be also the eigenstate of the $\vec{S}^{2}$ and $S_{z}$. In
Table \ref{tab:tab1}, the same values of ground state energy for
$d=0$ and $t=0$ system reflect SU(2) pseudospin symmetry.

The eigenvalues of $S_z$ are given by
$(N_{\uparrow}-N_{\downarrow})/2$. We find that our
charging-energy term is related to $S_z$ \cite{MacDonald} as
\begin{equation*}
-\frac{e^{2}}{\epsilon
l_{B}}\frac{d}{l_{B}}\frac{N_{\uparrow}N_{\downarrow}}{N}=\frac{e^{2}}{\epsilon
l_{B}}\frac{d}{l_{B}}\frac{-(N_{\uparrow}+N_{\downarrow})^{2}+(N_{\uparrow}-N_{\downarrow})^{2}}{4N}
=-\frac{e^{2}}{\epsilon l_{B}}\frac{d}{l_{B}}\frac{1}{4}N +
\frac{e^{2}}{\epsilon l_{B}}\frac{d}{l_{B}}\frac{1}{N}S_{z}^{2},
\end{equation*}
where the ground state prefers $S_z = 0$.

\begin{table}
\begin{ruledtabular}
\begin{tabular}{c|ccccc}
$(N_{\uparrow},N_{\downarrow})$     &  (4,4)   & (5,3)  & (6,2)  &  (7,1)  &  (8,0) \\
\hline
$H_{\uparrow}+H_{\downarrow}+H_{\uparrow\downarrow}$  & $-2.864848$  & $-2.864848$  &  $-2.864848$ &  $-2.864848$  &  $-2.864848$ \\
$-\frac{e^{2}}{\epsilon l_{B}}\frac{d}{l_{B}}\frac{N_{\uparrow}N_{\downarrow}}{N}$    & 0          & 0          &  0          &   0         &  0     \\
$H_{i}$ at $d / l_B =0.0$      &  $-2.864848$  & $-2.864848$  & $-2.864848$  & $-2.864848$  &  $-2.864848$  \\
\hline
$H_{\uparrow}+H_{\downarrow}+H_{\uparrow\downarrow}$  & $-2.678141$  & $-2.689105$  &  $-2.724817$  &  $-2.783162$   &  $-2.864848$  \\
$-\frac{e^{2}}{\epsilon l_{B}}\frac{d}{l_{B}}\frac{N_{\uparrow}N_{\downarrow}}{N}$   & $-0.2$          & $-0.1875$   & $-0.15$     & $-0.0875$      &  0 \\
$H_{i}$ at $d / l_B =0.1$    &  $-2.878141$  & $-2.876605$  &
$-2.874817$  &  $-2.870662$  &  $-2.864848$  \\
\end{tabular}
\end{ruledtabular}
\caption{Consistency checks for ground state energy. For $t=0.0$,
we can divide the Hilbert space according to $N_{\uparrow}$ and
$N_{\downarrow}$. We present the ground state energy in each
sector for the case of $t=0.0$ and $N=8$, comparing $d / l_B =0.0$
to $d / l_B =0.1$. Without the term of charging-energy, the sector
of $(N_{\uparrow},N_{\downarrow})=(8,0)$ contains the true ground
state at $d / l_B =0.1$ like Hund's rule.} \label{tab:tab1}
\end{table}

\section{Topological Entanglement Entropy}

In order to find the entanglement entropy which is a nonlocal
quantity \cite{Berrada}, we first conduct bipartition. We can
arbitrarily divide the space into two parts, $A$ and $B$. Then, we
calculate the von Neumann entropy $S_{A}$ as follows:
\begin{equation}
S_{A} = -\mbox{Tr}\rho_A \log_2 \rho_A, ~~~~~\rho_A = \mbox{Tr}_B
|\Psi_0 \rangle \langle \Psi_0 | .
\end{equation}
Here, the ground state $|\Psi_{0}\rangle$ with $M$ electrons is
spanned by the basis $\{ | i \rangle \}$:
\begin{eqnarray}
|\Psi_0 \rangle &=& \sum_{i}a_{i} | i \rangle \nonumber \\
&=& \sum_{i}a_{i} c^{\dagger}_{i_1} c^{\dagger}_{i_2}\cdots
c^{\dagger}_{i_{M}}|0 \rangle \nonumber \\
&=& \sum_{i}a_{i} (-1)^P c^{\dagger}_{j_1}\cdots
c^{\dagger}_{j_{M_A}} c^{\dagger}_{k_1} \cdots
c^{\dagger}_{k_{M_B}}|0 \rangle ,
\end{eqnarray}
where the creation operators should be reordered such that $j_1 <
\cdots < j_{M_A}$ in the subsystem $A$, $k_1 < \cdots < k_{M_B}$
in $B$, and $M_A + M_B = M$. The sign $(-1)^P$ should be taken
care of during the reordering step. In some systems, this entropy
$S_{A}$ indicates critical points of quantum phase transitions
\cite{Chung2}. However, this single entanglement entropy $S_{A}$
does not always indicate critical points. According to our
calculations, in bilayer quantum Hall systems, the single
entanglement entropy $S_{A}$ for bipartition of up layer $A$ and
down layer $B$ can not indicate critical points. Thus, it is
necessary to introduce slightly more complicated entropies in
addition to $S_{A}$ to form the order parameter. It is also
interest to investigate entanglement spectrum \cite{Schliemann2}.

Now let us turn our attention to the recent proposal \cite{Kitaev}
of the topological entanglement entropy. Our system described by
the Hamiltonian of Eq. (1) looks like a one-dimensional system
with a single index, because of the lowest Landau level
approximation. However, our original physical space is
two-dimensional and our system shows topological properties.

Since systems have the same number of quantum numbers as their
spatial dimensions, our system in two spatial dimensions has two
quantum numbers $(n,j)$ by ignoring spin and pseudospin, where $n$
denotes the Landau level, $n = 0 \cdots \infty $, and $j$ denotes
the intra-Landau level, $j=0\cdots N-1$. Here the Landau level
degeneracy $N$ plays the role of width in this two-dimensional
quantum number space. In the lowest Landau level approximation, we
restrict the quantum number space into $n=0$ space alone. Now we
should find a good bipartition of the lowest Landau level, where
the boundary length between $A$ and $B$ is proportional to $N$ to
guarantee large subspaces $A$ and $B$. The leading term of the
corresponding von Neumann entropy $S_{A}$ then can be proportional
to $N$ such as $S_{A} = \alpha N - \gamma +\cdots$. The even-odd
bipartition \cite{Chen} in the one-dimensional spin system was
considered to obtain a large entanglement entropy. We will apply
this kind of partition to our bilayer quantum Hall system.

In order to study a topological property of bilayer quantum Hall
systems with the Hamiltonian Eq. (1), we consider the topological
entanglement entropy $S_{topo}$ \cite{Kitaev} defined as
\begin{equation}
S_{topo} = S_{A} + S_{B} + S_{C} - S_{AB} - S_{BC} - S_{CA} +
S_{ABC},
\end{equation}
where it is crucial to partite the one-dimensional index system
into four lattice subsystems, $A$, $B$, $C$ and $D$. Keeping in
mind the requirement that all seven entropies in Eq. (6) should be
proportional to $N$, we make the following partition. As shown in
Fig. 1, we partite the index system of $\{j, \sigma \}$ in
$c^{\dagger}_{j\sigma}$ such as $j \mod 3 =0, 1, 2$ in up layer
($\sigma = \uparrow$) for $A$, $B$, $C$, respectively, and $D$
denotes all $j$ in down layer ($\sigma = \downarrow$). The
subsystem $AB$ represents $A \cup B$, etc. Numerical results
confirm that all entanglement entropies in Eq. (6) are
proportional to $N$ for small $d$.

\begin{figure}
\includegraphics[width= 9cm]{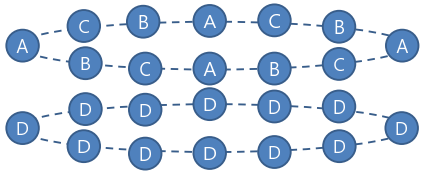}
%\vspace{13cm}
\caption {For $N=12$, a highly entangled partition into four
regions. This type of partition is adopted during the
calculations.} \label{fig:fig1}
\end{figure}

A slight change of topology does not effect on the value of
$S_{topo}$. However, it should be emphasized that a topologically
different partition provides a different entanglement entropy. For
example, we can divide the index system into four regions such as
$A'$ is even in up layer, $B'$ is odd in up layer, $C'$ is even in
down layer, and $D'$ is odd in down layer. We find however that
$S'_{topo}$ for this type of partition does not play the role of
the order parameter to explain the phase transition. In
consequence, we should make an adequate choice of partition to get
the proper topological entanglement entropy explaining the phase
transition.

\section{Numerical Works}

To compute the topological entanglement entropy $S_{topo}$, we
first look for the ground state $|\Psi_0 \rangle$ of the system,
by diagonalizing the Hamiltonian $H$ of Eq. (1). The main
calculation is to find the coefficients $a_{i}$ of Eq. (5) as the
coupling parameter $d$ changes for fixed $t$. The explicit forms
of the density matrices are determined by reordering indices in
Eq. (5). Then, we find the von Neumann entropies by diagonalizing
the density matrices. Finally the topological entanglement entropy
is obtained as the combination of the von Neumann entropies in Eq.
(6).

Before we present numerical results, the number of electrons $M$
is worthy of mentioning. Since the last term of the Hamiltonian in
Eq. (2) prefers equal number of electrons in up and down layers,
the ground state with even number of electrons has very small
electron number fluctuation in a typical finite size calculation.
Since the number fluctuation is essential for phase transition, we
should choose finite systems with odd number of electrons. Odd
number systems can circumvent unwanted effects of charging-energy
cost in finite systems because one additional electron fluctuating
between two layers does not cost charging-energy.

Because of the advantage of odd number of electrons, we have
computed the topological entanglement entropy $S_{topo}$ up to odd
$M=13$ as a function of the layer separation $d$ for various
values of the tunneling amplitude $t$, which is plotted in Fig. 2.
The shape of the topological entanglement entropy in bilayer
quantum Hall systems is similar to the spontaneous magnetization
in the Ising model \cite{Landau}.

\begin{figure*}[ht]
\centering
\includegraphics[width= 6.0 cm]{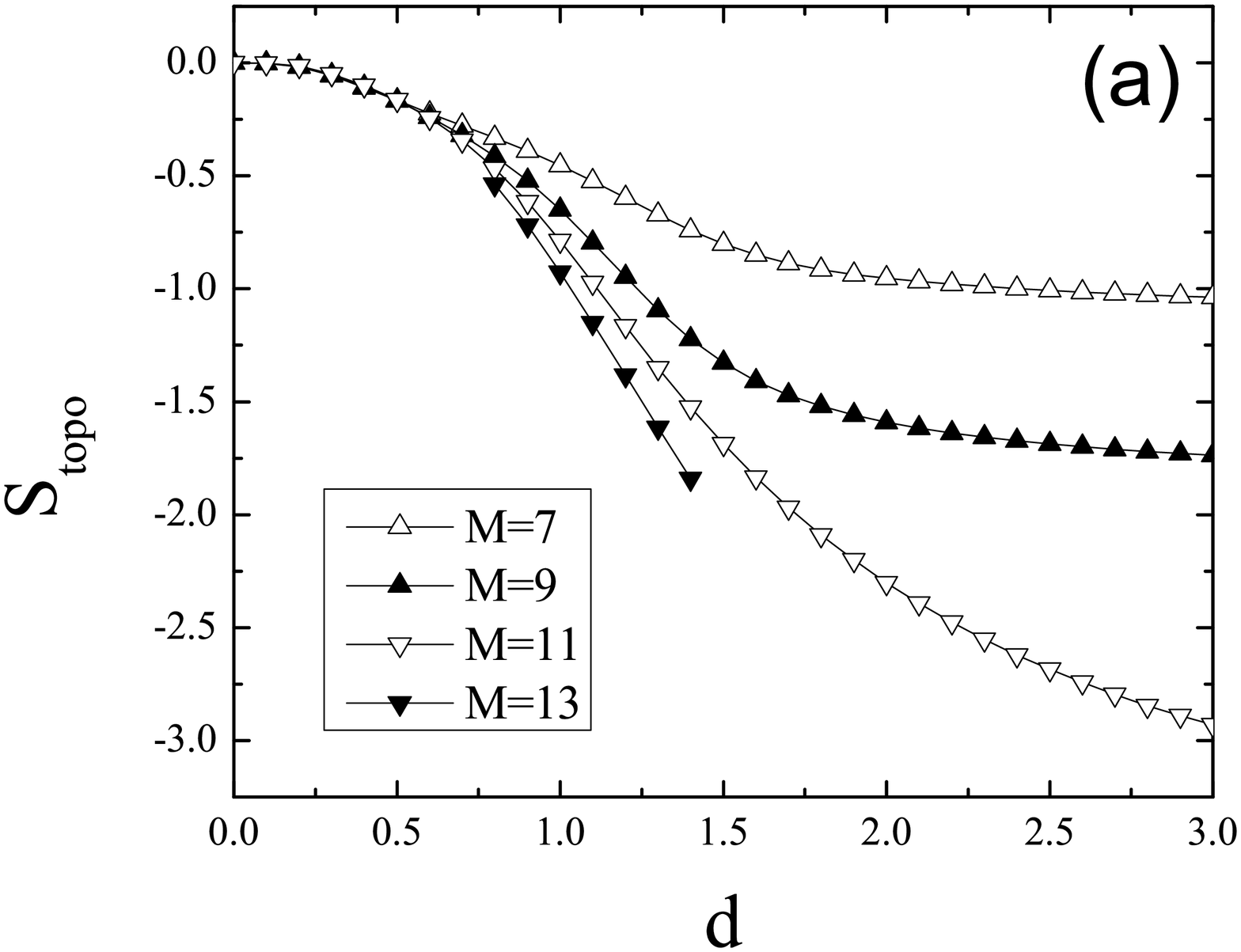}
\includegraphics[width= 6.0 cm]{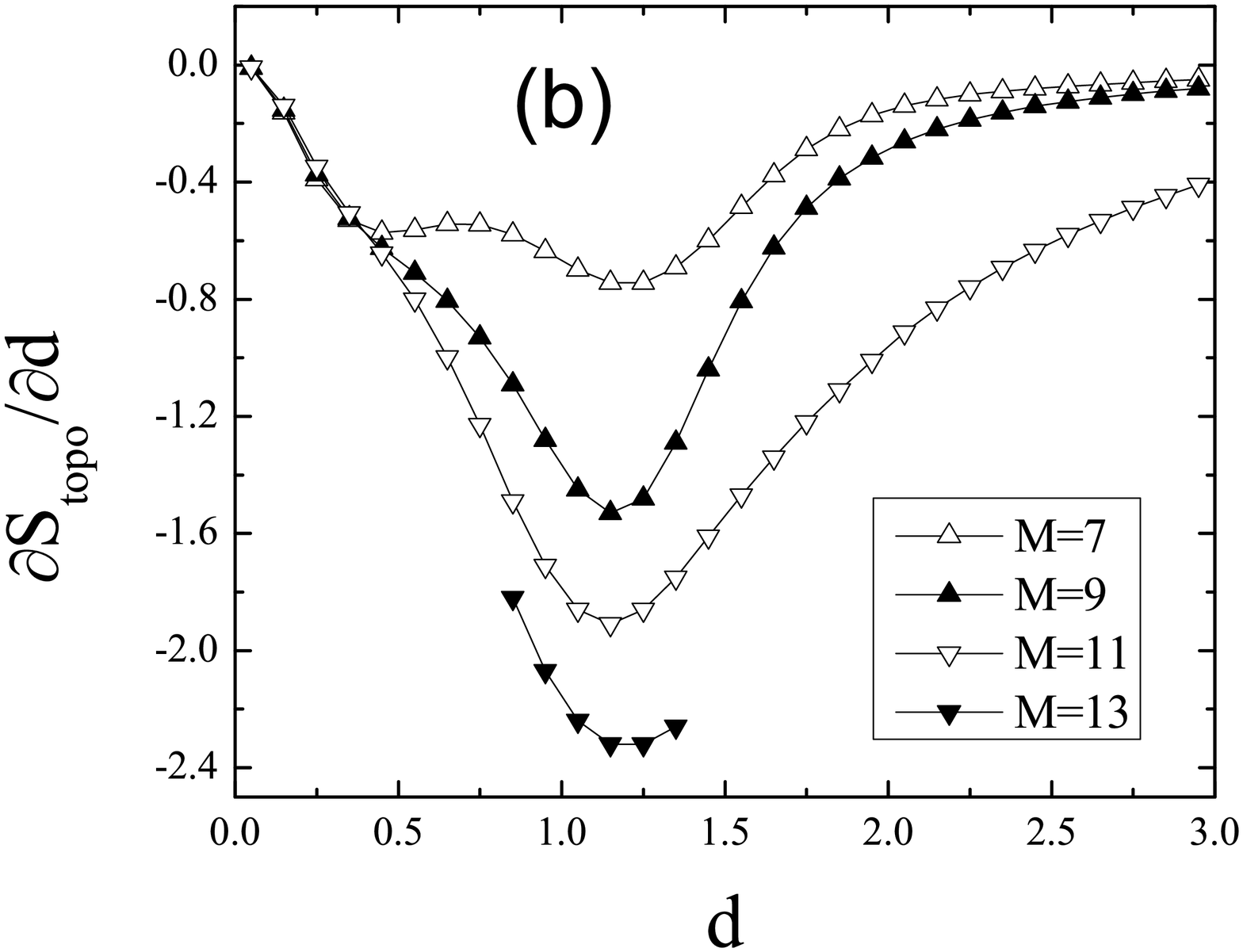}
\includegraphics[width= 6.0 cm]{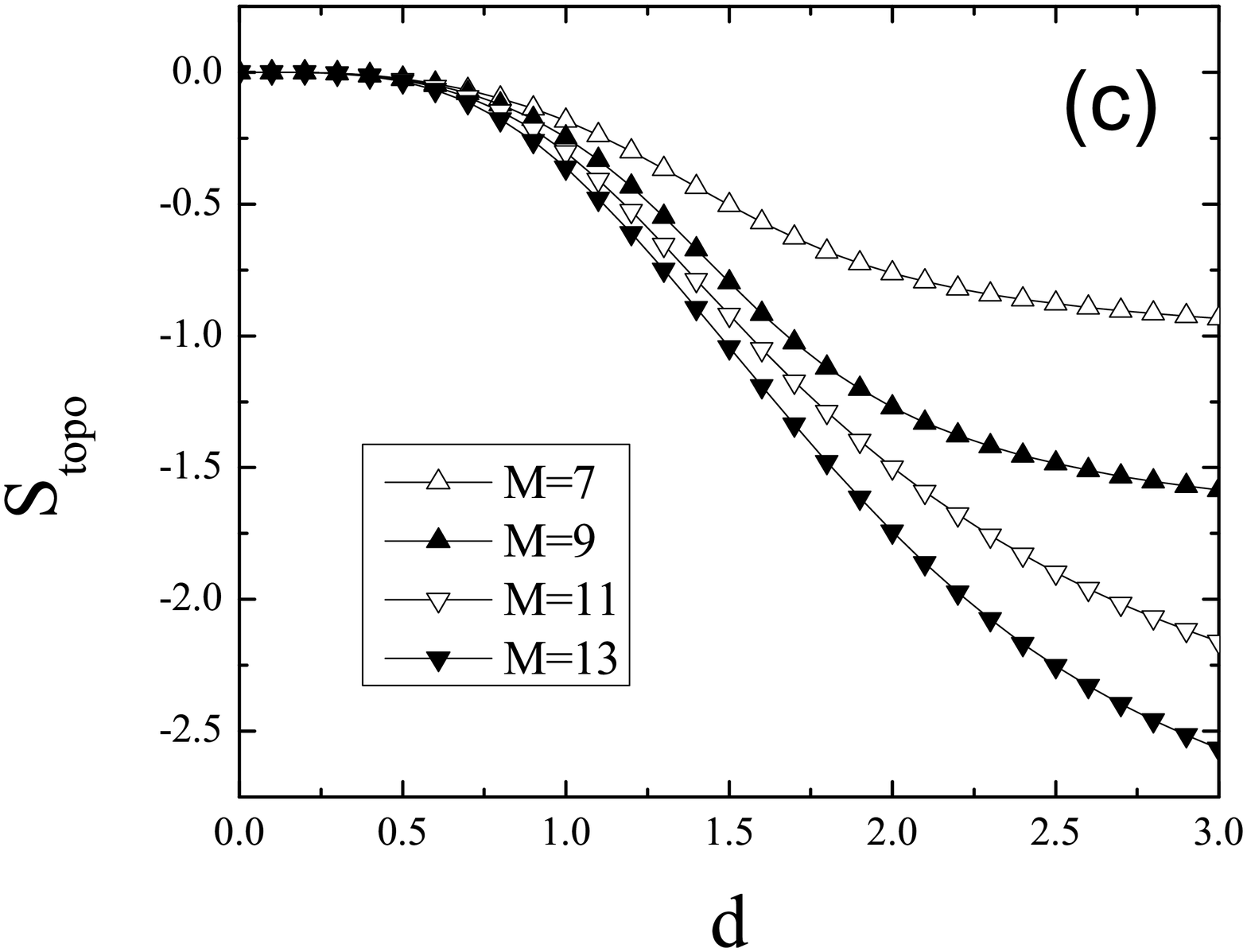}
\includegraphics[width= 6.0 cm]{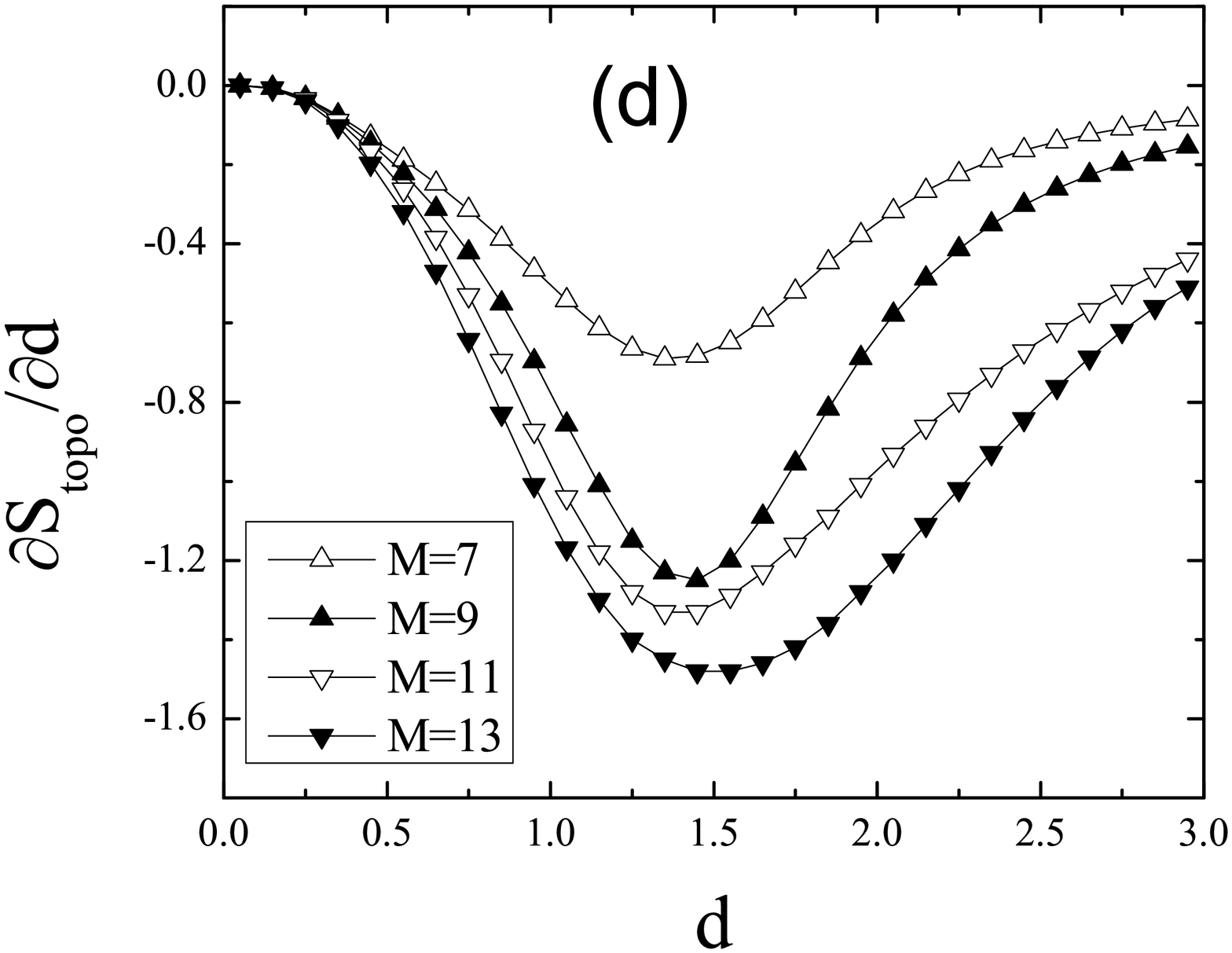}
\includegraphics[width= 6.0 cm]{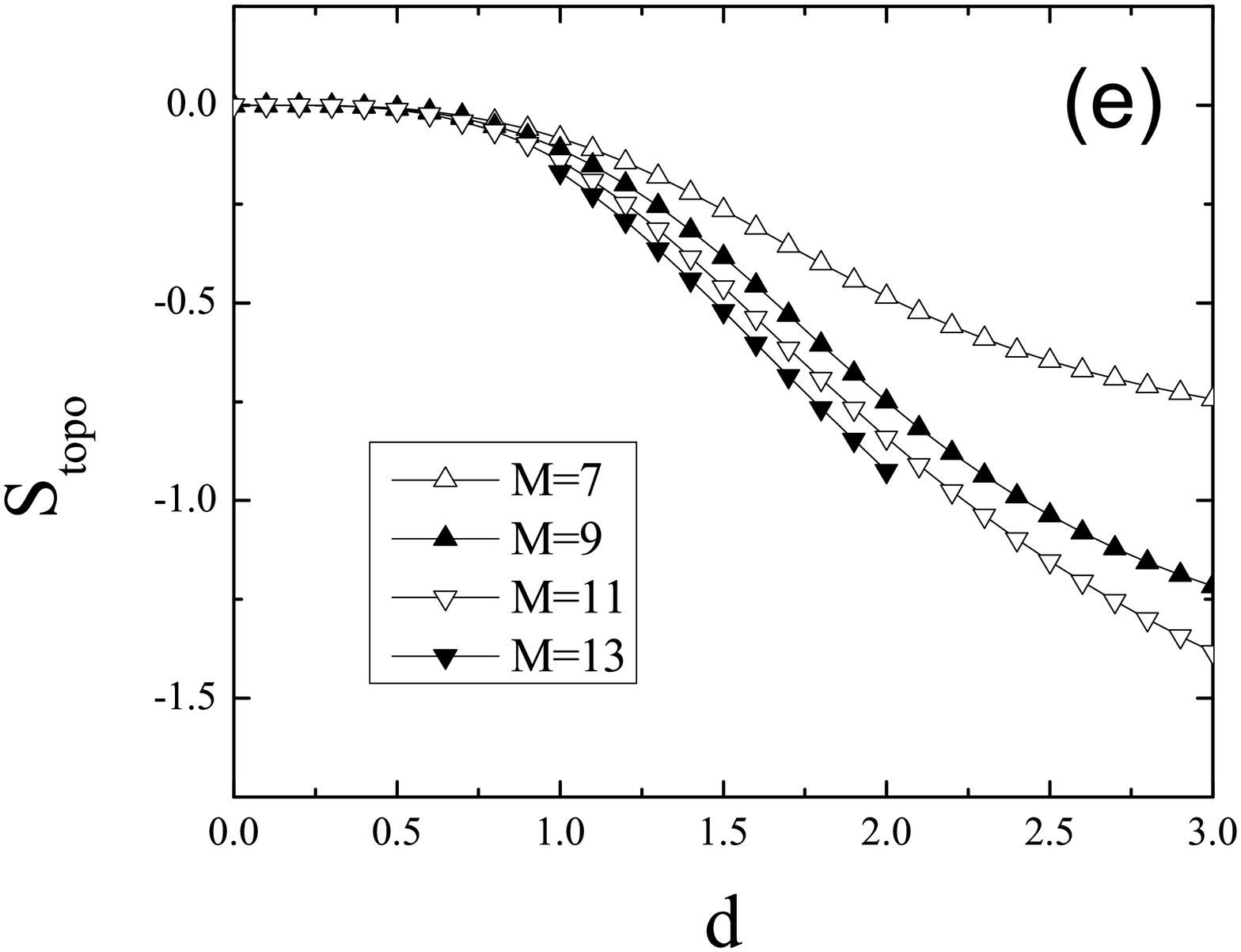}
\includegraphics[width= 6.0 cm]{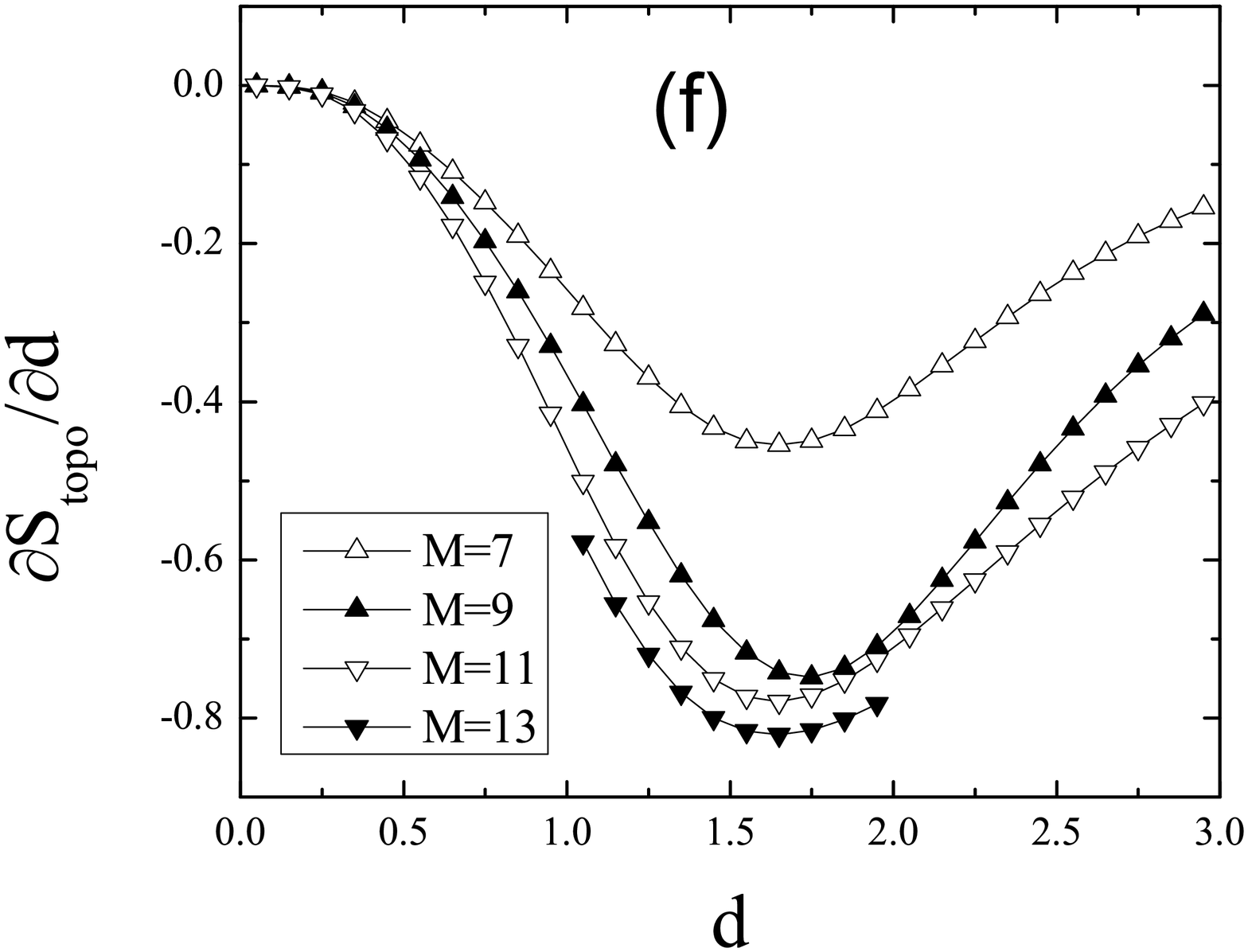}
\caption {The entropy $S_{topo}$ versus the layer separation $d$,
and the derivative of the entropy $\partial S_{topo} /
\partial d$ versus $d$ for three fixed values of $t$:
(a) and (b) $t=0.01$, (c) and (d) $t=0.05$, (e) and (f) $t=0.1$.
We notice that there are the pronounced dips in $\partial S_{topo}
/ \partial d$ at nearly the same $d$ for each $t$.}
\label{fig:fig2}
\end{figure*}

\begin{figure*}[ht]
\centering
\includegraphics[width= 6.0 cm]{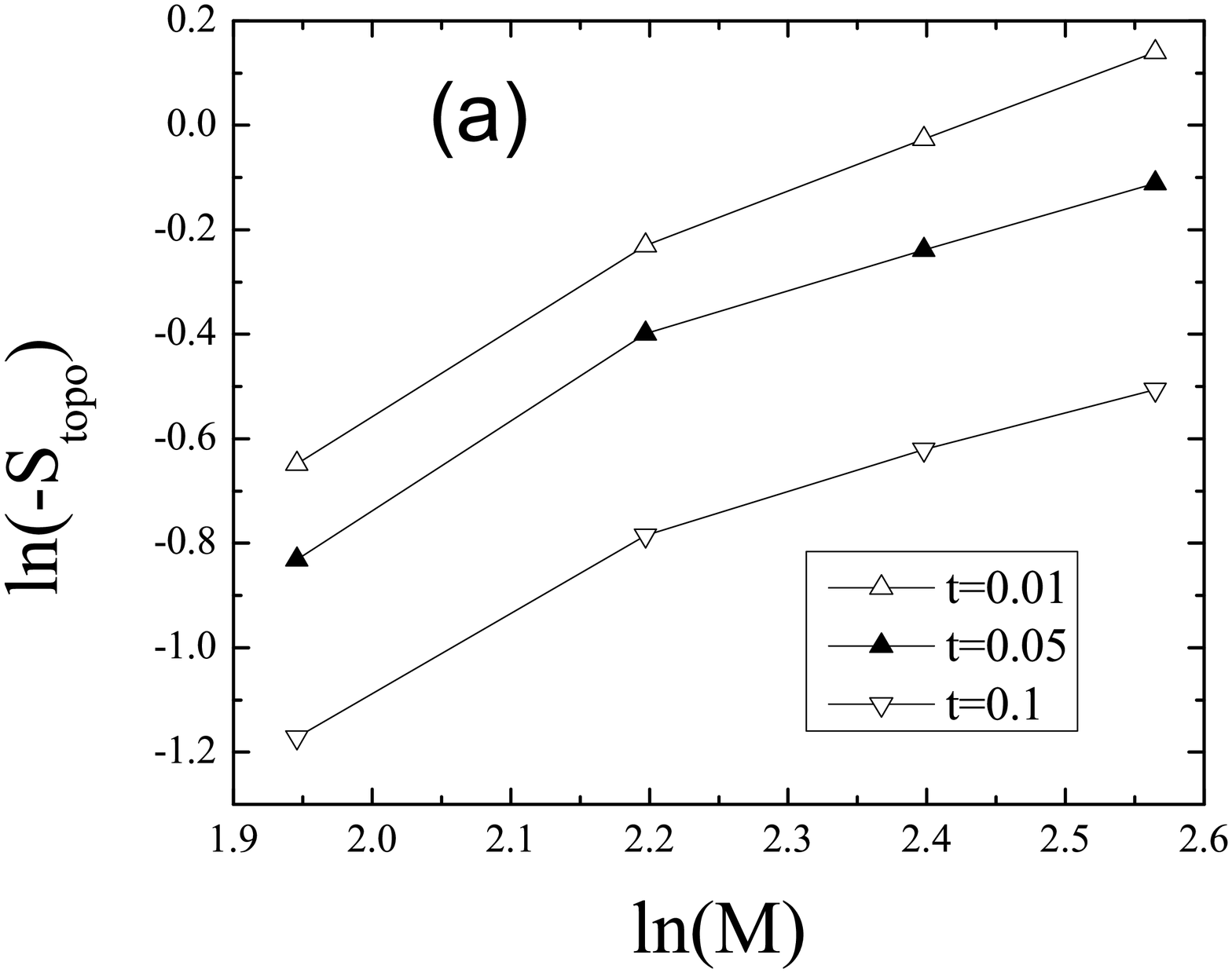}
\includegraphics[width= 6.0 cm]{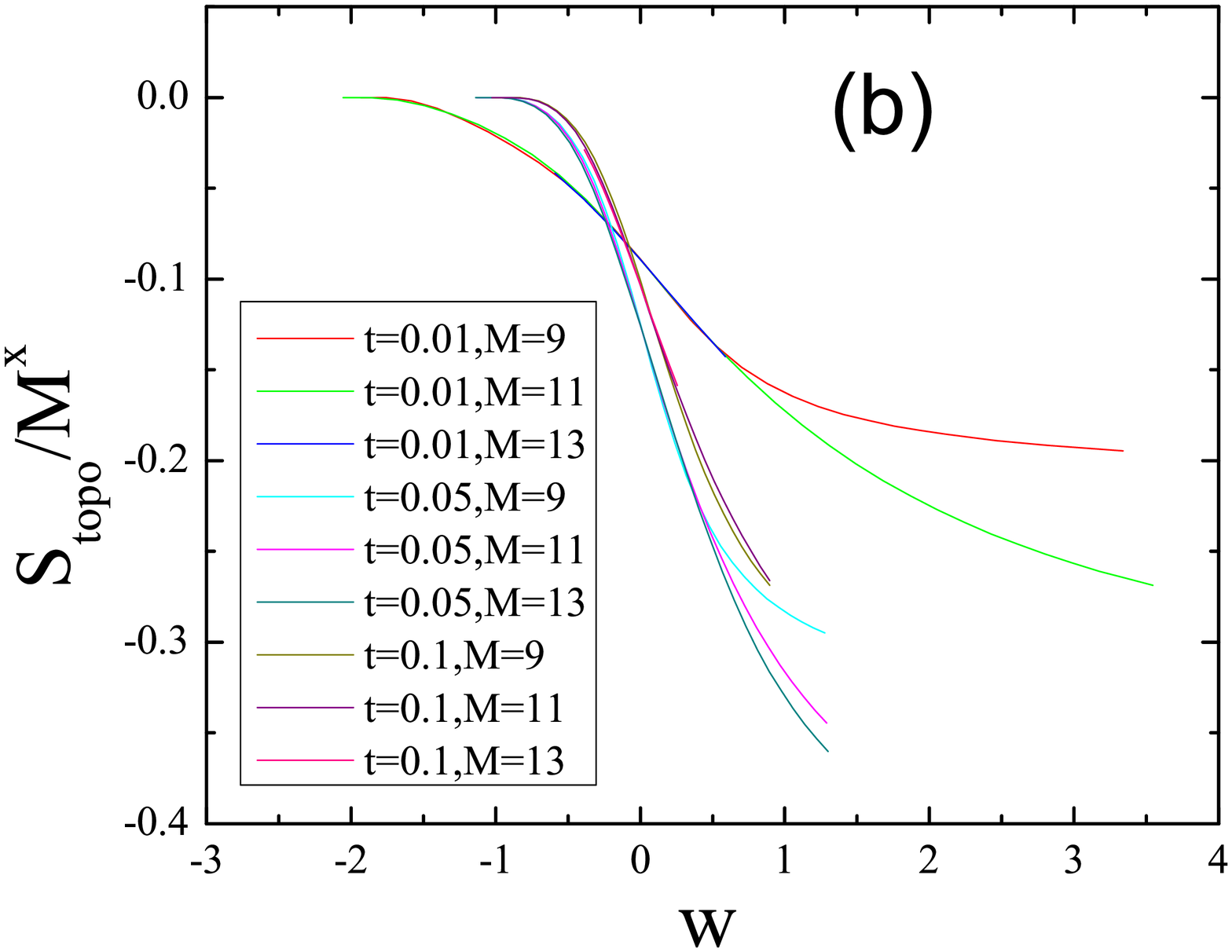}
\caption {(Color online) (a) The logarithm of the entropy
$\ln(-S_{topo})$ versus the logarithm of the number of electrons
$\ln(M)$. Ignoring the data of $M=7$, we notice the linear
relationship between the two values. It is easy to determine the
slope which gives $x$ shown in Table \ref{tab:tab2}. (b) The
entropy $S_{topo}$ versus the scaled parameter $w=\epsilon
M^{1/\nu}$ where $\epsilon = (d - d_c )/d_c$. A good data collapse
is found at $1/\nu = 0.3$ for the data of $t=0.01$, at $1/\nu =
0.05$ for the data of $t=0.05$, and at $1/\nu = 0.01$ for the data
of $t=0.1$.} \label{fig:fig3}
\end{figure*}

In general, the phase boundary between quantum Hall state and
compressible states can be obtained at the point where the
topological entanglement entropy changes most and it has the
maximum negative slope. Fig. 2 shows the maximum negative slopes.
Specifically, the plot of $\partial S_{topo} /
\partial d$ in Fig. 2(b) shows a dip at $d \simeq 1.1 l_B$ for
$t=0.01$. This may be a signature of transition from coherent to
incoherent quantum phase.

We postulate the scaling function $\Theta$ to present the
topological entanglement entropy $S_{topo}$ as
\begin{equation}
S_{topo} = M^x \Theta (\epsilon M^{\frac{1}{\nu}}).
\end{equation}
For the purpose of making a data-collapse, we look for the scaling
exponents $x$ and $1/\nu$ as a function of the tunneling amplitude
$t$. In order to find the exponent $x$, we use the data of the
critical point where $\epsilon \equiv (d - d_c )/ d_c = 0$. In
Fig. 3(a), we find that $S_{topo}$ is very well proportional to
$M^x$ as far as we ignore the data of $M=7$, that is, $\ln M
\simeq 1.946$. Once we have determined $d_c$ and $x$, the
determination of $1/\nu$ gives the data-collapse. As shown in Fig.
3(b), all data of Fig. 2 are collapsed into a single line near at
the scaled parameter $w \equiv \epsilon M^{\frac{1}{\nu}}=0$. The
numerical results are summarized in Table \ref{tab:tab2}.

\begin{table}
\begin{ruledtabular}
\begin{tabular}{c|ccc}
$t$      & $d_c$  & $x$      &  $1/\nu$\\
\hline
$0.01$   & $1.1 \pm 0.04$  & $0.99 \pm 0.03$ &  $0.3  \pm 0.05$  \\
$0.05$   & $1.4 \pm 0.05$  & $0.76 \pm 0.05$ &  $0.05 \pm 0.1$ \\
$0.1$    & $1.6 \pm 0.06$  & $0.68 \pm 0.08$ &  $0.01 \pm 0.1$ \\
\end{tabular}
\end{ruledtabular}
\caption{\label{tab:1}Summary of numerical results. For a given t,
the values of $d_c$, $x$, and $1/\nu$ are calculated. For
$t=0.01$, the data-collapse at $1/\nu = 0.3$ looks more likely a
single line than at $1/\nu = 0.35$, so that we roughly estimate
the error. However, it is hard to distinguish the better
data-collapse by changing the value of $1/\nu$ for $t=0.05$ or
$t=0.1$, so that we present a bigger error.} \label{tab:tab2}
\end{table}

To determine the phase boundary between coherent and incoherent
states, we use the first derivative of topological entanglement
entropy with respect to $d$. Note that the maximum of the
first-derivative is a function of $t$. We drew the phase boundary
by plotting the maxima of the first-derivatives with respect to
$d$ in Fig. 4. The phase boundary has a slight increasing tendency
as the tunneling amplitude increases. The phase boundary line
looks concave from below instead of convex. The parabolic (convex)
phase boundary was proposed by Murphy {\it et al.} \cite{Murphy}.
However, the parabolic phase boundary has a little discrepancy
\cite{Spielman2} with the tunneling data as shown in Fig. 5. In
fact, the critical value $d_{c}$ at $t=0$ determined by the
tunneling data is less than the value given by the parabolic phase
boundary line. Another point of the parabolic phase boundary is
that there always exists an enough big tunneling amplitude $t$
that produces a quantum Hall state for any $d$. This means that
the tunneling can produce a quantum Hall state without the
inter-layer Coulomb interaction for infinitely separated bilayer
systems. This dominant role of tunneling may be overestimated to
produce a quantum Hall state. In order to explain the transport
data and the tunneling data simultaneously, we modify the phase
boundary line which is concave as shown in Fig. 5. The tunneling
enhances coherence, but the inter-layer Coulomb interaction is
more important for a quantum Hall state. The concave phase
boundary may be reasonable.

It is known that the phase boundary line in Fig. 4 shifts upward
for the system with finite layer thickness \cite{Shibata2}.
Although we do not calculate explicitly here, finite layer
thickness will be crucial when we try to adjust the critical
values of $d_{c}$.

\begin{figure}
\includegraphics[width= 9cm]{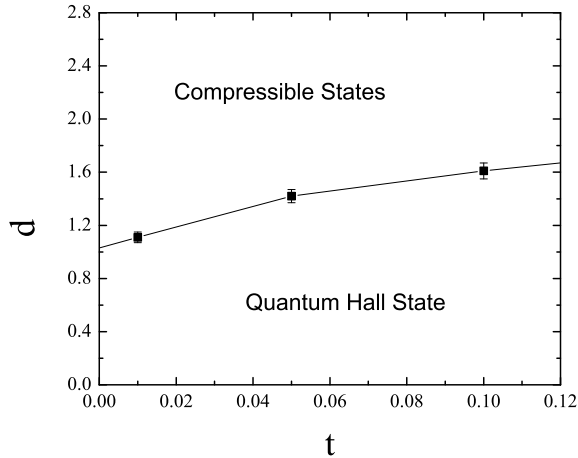}
%\vspace{13cm}
\caption {The phase boundary in the parameter space of $d$ and
$t$. Slightly the less steep slopes are observed as $t$
increases.} \label{fig:fig4}
\end{figure}

\begin{figure}
\includegraphics[width= 9cm]{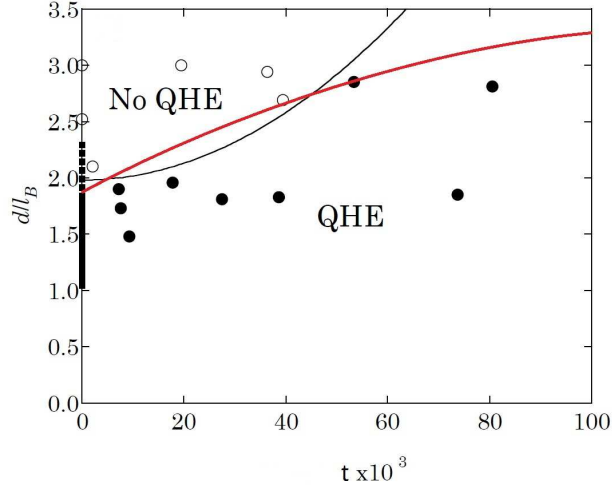}
%\vspace{13cm}
\caption {Data presented in Spielman's thesis. Each circle
represents a sample as measured by magneto-transport. Solid
markers indicate the existence of a quantum Hall minimum, and open
markers the lack thereof. The bold lines on the vertical axis
represent tunneling data: the solid portion indicates the
existence of a peak in tunneling and the dashed portion indicates
its absence. The black line is a proposed parabolic phase boundary
based on the transport data by Murphy {\it et al.} However, it is
possible to propose a modified phase boundary line, which is
concave from below like the red line. This red line is consistent
with the tunneling data.} \label{fig:fig5}
\end{figure}

\section{Conclusion}

We have considered the quantum phase transition controlled by the
layer separation in bilayer quantum Hall systems. The interaction
between electrons in the system is described by the Coulomb
interaction in a torus geometry within the lowest Landau level
approximation. The main numerical work is to compute the
topological entanglement entropy by exact diagonalization. We find
that the topological entanglement entropy plays the role of an
order parameter to distinguish quantum phases.

In summary, we have presented the topological entanglement entropy
in bilayer quantum Hall systems. We have concluded that the
topological entanglement entropy is a better order parameter than
the pseudospin magnetization in bilayer quantum Hall systems. The
quantum order in bilayer quantum Hall systems is originated by
topological properties.

\begin{acknowledgements}

This work was partially supported by Basic Science Research
Program through the National Research Foundation of Korea(NRF)
funded by the Ministry of Education, Science and Technology(Grant
No. 2011-0023395), and by the Supercomputing Center/Korea
Institute of Science and Technology Information with
supercomputing resources including technical support(Grant No.
KSC-2012-C1-09). The author would like to thank K. M. Choi, S. J.
Lee, and J. H. Yeo for helpful discussions. The author is grateful
to S. M. Girvin who suggested this research topic several years
ago.

\end{acknowledgements}

\end{document}